\documentclass[aps,prl,twocolumn,showpacs,superscriptaddress,groupedaddress]{revtex4-1}  
\usepackage{graphicx}  
\usepackage{dcolumn}   
\usepackage{bm}        
\usepackage{amssymb}   
\usepackage{amsmath}   

\allowdisplaybreaks

\begin{document}


\title{Tunneling Current Measurement Scheme to Detect 
 Majorana-Zero-Mode-Induced Crossed Andreev Reflection}

\author{Lei Fang$^{1,2}$}
\email{lfang@gradcenter.cuny.edu} 
\author{David Schmeltzer$^{1,2}$}
\author{Jian-Xin Zhu$^{3,4}$}
\author{Avadh Saxena$^{3}$}
\affiliation{$^{1}$Physics Department, City College of the CUNY, New York, New York 10031, USA} 
\affiliation{$^{2}$Graduate Center, CUNY, 365 5th Ave., New York, New York 10016, USA} 
\affiliation{$^{3}$Theoretical Division, Los Alamos National Laboratory, Los Alamos, New Mexico 87545, USA}
\affiliation{$^{4}$Center for Integrated Nanotechnologies, Los Alamos National Laboratory, Los Alamos, New Mexico 87545, USA}

\begin{abstract}
We propose a scheme to detect the Majorana-zero-mode-induced crossed Andreev reflection by measuring tunneling current directly. In this scheme a metallic ring structure is utilized to separate electron and hole signals. Since tunneling electrons and holes have different propagating wave vectors, the conditions for them to be constructively coherent in the ring differ. We find that when the magnetic flux threading the ring varies, it is possible to observe adjacent positive and negative current peaks of almost equal amplitudes.  
\end{abstract}

\maketitle

Majorana zero modes (MZM) have been attracting the interest of the condensed matter physics community in recent years~\cite{Alicea_Review} due to their potential application in topological quantum computation~\cite{Nayak_TQC_Review}.  Several proposals have been made to realize MZM in realistic systems~\cite{Read_Green, FuKane_MF_TIsurf, Sau_MF_SOC_Zeeman_sWave}. Among them, the last one~\cite{Sau_MF_SOC_Zeeman_sWave} is based on the structure of a strong spin-orbit coupled semiconductor nanowire in proximity to an s-wave superconductor. This proposal, due to its convenience, has spurred many experimental efforts immediately~\cite{Mourik_Ex_MF_Semi_SC_Wire, Das_Ex_ZBP, Deng_Ex_ZBP, Churchill_Ex_ZB_oscillation}. More recently, several other different methods~\cite{Nadj_Ex_MZM_Fe_Chain,JPXu_Ex_MZM_Vortex_TSC,QLHe_Ex_ChiralMF_QSH} have been applied for the observation of MZM.

Several methods can be used to detect the existence of MZM~\cite{Alicea_Review}, and the tunneling spectroscopy is one of them. It is generally based on the MZM-induced Andreev reflection (AR)~\cite{Nilsson_CAR, Law_MF_Induced_AR}. There are two types of AR: local AR (LAR) and crossed AR (CAR). MZM-induced LAR results in a zero-bias peak (ZBP) of the differential conductance in the tunneling process. Up until now, ZBP has been observed by many experimental groups~\cite{Mourik_Ex_MF_Semi_SC_Wire, Das_Ex_ZBP, Deng_Ex_ZBP}. However, the conclusion of the existence of MZM cannot be made unambiguously through the observation of ZBP only, for some other mechanisms may produce similar results. More experimental evidence is needed to obtain definitive results. In contrast to LAR, it is believed that MZM-induced CAR does not result in any charge tunneling signal, since the electron tunneling probability equals to that for the hole in CAR. Instead of the tunneling current, it is suggested that shot noise~\cite{Blanter_Buttiker_ShotNoise_Review} has special properties in MZM-induced CAR and can, in principle, be detected. Nonetheless, to the best of our knowledge, no experiment has ever been done till now to measure the shot noise in a CAR set-up. 

In this work, we propose a new scheme to observe the tunneling current directly through the MZM-induced CAR process. It goes beyond the traditional belief that the shot noise measurement is necessary in the observation of MZM-induced CAR. In order to separate the electron and hole tunneling signals, we employ a metallic ring to hold tunneling electrons and holes. A magnetic flux is threaded through the ring, so that an electron/hole acquires an Aharonov-Bohm (AB) phase when traveling around the ring~\cite{AB_Paper}. The resonance condition for the electron/hole to be constructively coherent in the ring is 
\begin{equation} \label{eq:Resonant_Condition}
\phi_D \pm \phi_M = 2n\pi  \quad  (n \in \mathbb{Z}) ,
\end{equation}
where $\phi_D = kL$ ($k$ is the wave vector of the electron/hole, $L$ is the circumference of the ring) and $\phi_M = 2\pi \Phi / \Phi_0$ ($\Phi$ is the magnetic flux threading the ring in units of the flux quantum $\Phi_0$) are the dynamical phase and the magnetic phase (AB phase) that an electron/hole acquires when traveling one complete circle around the ring, and $\pm$ denotes clockwise/counterclockwise circulations. Because under a finite bias, tunneling electrons (above the Fermi level) have slightly larger wave vector than tunneling holes (below the Fermi level), they are not constructively coherent simultaneously.  If we couple an external lead weakly to the ring, only when the resonant condition is fulfilled, there is a remarkable tunneling signal. Therefore, one can expect to observe electron and hole tunneling peaks separately corresponding to different magnetic fluxes.

\begin{figure}  [t]
\begin{center}
\includegraphics[scale=0.28]{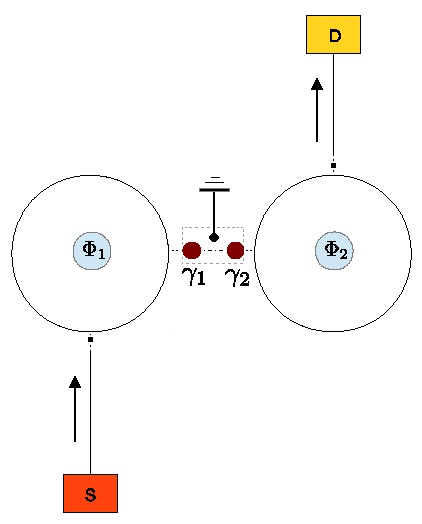}
\end{center}
\caption{A schematic setup to measure MZM-induced crossed Andreev reflection. Here $\gamma_1$ and $\gamma_2$ denote the two coupled MZM. Two metallic rings, threaded by magnetic fluxes $\Phi_1$ and $\Phi_2$, are coupled to the two MZM and two external leads through the tunneling effect. The superconductor hosting the MZM is grounded. Electrons are incident from the lead contacting to the source (S). Tunneling electrons/holes outflow along the second lead to the drain (D).} \label{fig:Transport_Configuration}
\end{figure}

Suppose there exists a pair of MZM close in space with a coupling energy $E_M$. The pair can be the MZM located at the opposite ends of a topological superconducting wire~\cite{Sau_MF_SOC_Zeeman_sWave}. The Hamiltonian for the two coupled MZM is 
\begin{equation} \label{eq:Hamiltonian_Majorana_Overlapping}
H_M = i\frac{E_M}{2}\gamma_1\gamma_2  \; .
\end{equation}
Here $\gamma_1$ and $\gamma_2$ are Majorana operators that satisfy $\gamma_1^\dagger = \gamma_1$, $\gamma_2^\dagger = \gamma_2$, $\gamma_1^2=\gamma_2^2=1$ and $\gamma_1\gamma_2 + \gamma_2\gamma_1 = 0$. Then, the MZM $\gamma_i$ ($i=1,2$) is coupled to the $i$-th ring through a tunneling Hamiltonian
\begin{equation} \label{eq:Hamiltonian_Tunnelling}
H^{(i)}_T = \int dx \frac{t_i}{\sqrt{2}} \gamma_i [\psi_i^\dagger(x) - \psi_i(x)]\delta(x) \; ,
\end{equation} 
where $\psi_i(x)$ and $\psi_i^\dagger(x)$ are electron annihilation and creation operators in the $i$-th ring, which satisfy the usual fermion commutation relations, and $t_i$ denotes the coupling strength.  

\begin{figure}  [t]
\begin{center}
\includegraphics[scale=0.45]{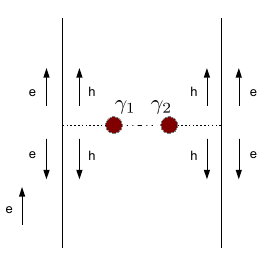}
\end{center}
\caption{Majorana-zero-modes-induced crossed Andreev reflection. An electron is incident from the lower side of the left metallic wire, which is coupled through a pair of Majorana zero modes to the right metallic wire. It can: (1) reflect with amplitude $r_1^{(e)}$; (2) transmit to the upper side with amplitude $t_1^{(e)}$; (3),(4) tunnel to either lower or upper side of the right wire with amplitude $w_1^{(e)}$; (5),(6) be converted to holes through local Andreev reflection and flow to either lower side with amplitude $r_1^{(h)}$ or upper side with amplitude $t_1^{(h)}$; (7),(8) be converted to holes through crossed Andreev reflection and flow to either lower or upper side of the right wire with amplitude $w_1^{(h)}$.} \label{fig:CAR}
\end{figure}

Our proposed transport experiment set-up is shown in Fig.~\ref{fig:Transport_Configuration}. We let one lead bridge a source reservoir to the left ring and another lead bridge the right ring to a drain reservoir.  The coupling between a lead and a ring can be described by a tri-junction scattering matrix~\cite{Tri-junction_Scattering}
\begin{equation} \label{eq:Tri_Junction_S-Matrix}
S_L = \left(
\begin{array}{ccc}
-(a+b) & \sqrt{\epsilon} & \sqrt{\epsilon} \\
\sqrt{\epsilon} &  a  &  b  \\
\sqrt{\epsilon} &  b  &  a
\end{array}
\right) \, ,
\end{equation}
where $\epsilon$ denotes the tunneling strength, $a=(\sqrt{1-2\epsilon}-1)/2$ and $b=(\sqrt{1-2\epsilon}+1)/2$. The lead and the two ring-arms consist of the three channels of this scattering matrix. First, electrons incident from the source flow along the left lead and tunnel into the left ring. Next, electrons in the left ring tunnel into the right ring (may be converted to holes at the same time) through MZM-induced CAR. This process is shown in Fig.~\ref{fig:CAR} and can be described by a scattering matrix 
\begin{equation} \label{eq:CAR_S-Matrix}
S_M = \left(
\begin{array}{cccccccc}
r_1^{(e)} & t_1^{(e)} & w_2^{(e)} & w_2^{(e)} & \tilde{r}_1^{(e)} & \tilde{t}_1^{(e)} & \tilde{w}_2^{(e)} & \tilde{w}_2^{(e)}   \\
t_1^{(e)} & r_1^{(e)} & w_2^{(e)} & w_2^{(e)} & \tilde{t}_1^{(e)} & \tilde{r}_1^{(e)} & \tilde{w}_2^{(e)} & \tilde{w}_2^{(e)}   \\
w_1^{(e)} & w_1^{(e)} & r_2^{(e)} & t_2^{(e)} & \tilde{w}_1^{(e)} & \tilde{w}_1^{(e)} & \tilde{r}_2^{(e)} & \tilde{t}_2^{(e)}   \\
w_1^{(e)} & w_1^{(e)} & t_2^{(e)} & r_2^{(e)} & \tilde{w}_1^{(e)} & \tilde{w}_1^{(e)} & \tilde{t}_2^{(e)} & \tilde{r}_2^{(e)}   \\
r_1^{(h)} & t_1^{(h)} & w_2^{(h)} & w_2^{(h)} & \tilde{r}_1^{(h)} & \tilde{t}_1^{(h)} & \tilde{w}_2^{(h)} & \tilde{w}_2^{(h)}   \\
t_1^{(h)} & r_1^{(h)} & w_2^{(h)} & w_2^{(h)} & \tilde{t}_1^{(h)} & \tilde{r}_1^{(h)} & \tilde{w}_2^{(h)} & \tilde{w}_2^{(h)}   \\
w_1^{(h)} & w_1^{(h)} & r_2^{(h)} & t_2^{(h)} & \tilde{w}_1^{(h)} & \tilde{w}_1^{(h)} & \tilde{r}_2^{(h)} & \tilde{t}_2^{(h)}   \\
w_1^{(h)} & w_1^{(h)} & t_2^{(h)} & r_2^{(h)} & \tilde{w}_1^{(h)} & \tilde{w}_1^{(h)} & \tilde{t}_2^{(h)} & \tilde{r}_2^{(h)}
\end{array}
\right) \  ,
\end{equation} 
of which the eight channels are lower left electron, upper left electron, lower right electron, upper right electron, lower left hole, upper left hole, lower right hole and upper right hole. For the detailed form and a derivation of these scattering matrix elements, see the supplementary material~\cite{Supplement_Material}. The MZM-induced CAR dominates LAR when the electron incident energy $E$ and the MZM energy width $t_i^2/\hslash v_F$ ($v_F$ is the Fermi velocity) are much smaller than the MZM energy $E_M$~\cite{Nilsson_CAR}. Lastly, electrons/holes in the right ring tunnel to the second lead, and outflow to the drain. 

\begin{figure}  [t]
\begin{center}
\includegraphics[scale=0.4]{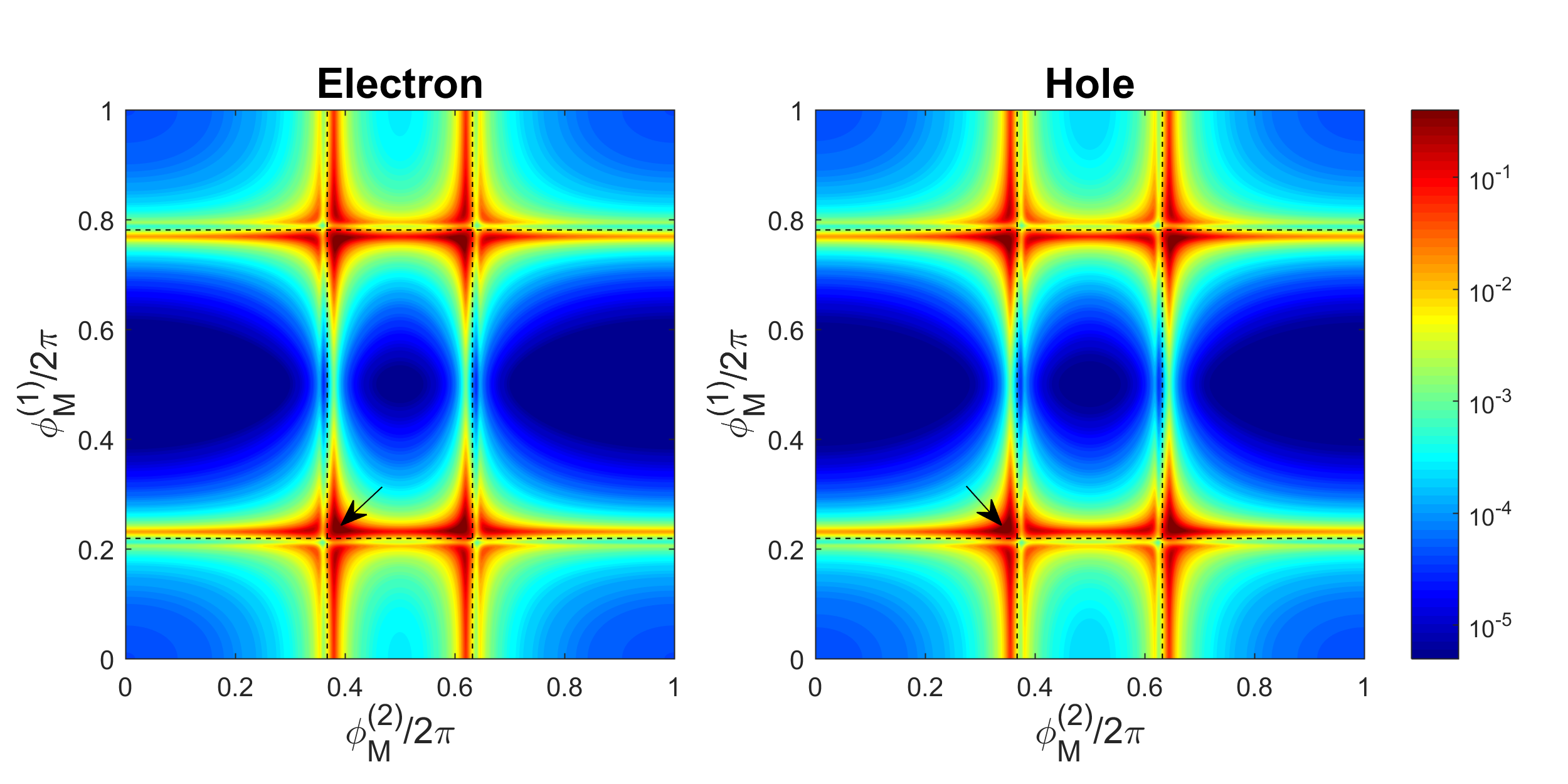}
\end{center}
\caption{``Heat map'' for the tunneling probabilities of electron (left) and hole (right). An electron is assumed to be incident from the first lead. The vertical (horizontal) axis represents the magnetic phase an electron/hole acquire when traveling one complete circle around the left (right) ring. A log-scale in color is utilized to represent the transmission probabilities. The parameters we set to make this plot are: Fermi vector $k_F=0.15~ \text{nm}^{-1}$, Fermi energy $E_F=12.835 ~\text{meV}$, electron incident energy $E=0.003 ~\text{meV}$, circumference of the left ring $L_1=2.69 ~\mu\text{m}$, circumference of the right ring $L_2=2.78 ~\mu\text{m}$, MZM coupling energy $E_M=0.01~\text{meV}$, MZM width $\Gamma=t^2/\hslash v_F = 0.001 ~\text{meV}$, scattering parameter at the lead-ring tri-junction $\epsilon=0.05$. Under this setting, we have the dynamical phase of the electron/hole circling the left ring $\phi_{e/h}^{(1)}=(64.219\pm0.008)\times 2\pi$ and the dynamical phase of the electron/hole circling the right ring $\phi_{e/h}^{(2)}=(66.368\pm0.008)\times 2\pi$. Dashed lines in the graph correspond to the magnetic phases that fulfill $k_F L_1 \pm \phi_M^{(1)}=2n\pi$ and $k_F L_2 \pm \phi_M^{(2)}=2n^\prime \pi$, which are $\phi_M^{(1)}=0.219 \times 2\pi, ~ 0.781 \times 2\pi$ and $\phi_M^{(2)}=0.368 \times 2\pi, ~ 0.632 \times 2\pi$. The arrows indicate the tunneling peaks (dark red) close to the magnetic phases $\phi_{M}^{(1)} = 0.219 \times 2\pi$, $\phi_{M}^{(2)} = 0.368 \times 2\pi$. The electron peak and the hole peak are located both above the line $\phi_{M}^{(1)} = 0.219 \times 2\pi$ because it is the electron (not hole) incident from the first lead, dominating in the left ring. The electron peak is to the right and the hole peak to the left of the line $\phi_{M}^{(2)} = 0.368 \times 2\pi$, due to that the dynamical phase of the tunneling electron/hole in the right ring is above/below $k_F L_2 = 66.368\times 2\pi$.} \label{fig:Transmission}
\end{figure} 
 
We require that the metallic rings are coupled weakly to both the leads and the MZM. In this case the rings are almost isolated, and the tunneling electrons/holes can stay in the ring for a relatively long time. Thus, the resonance condition Eq.(\ref{eq:Resonant_Condition}) plays an important role in the transport process.  
 
The function of the left ring is to single out the incident electron energy. Only when an electron is constructively coherent in the ring (i.e., fulfills the resonant condition), it has a substantial contribution to the incident flow of CAR. It is worth mentioning that two directly coupled metallic rings were investigated in previous works~\cite{David_Two_Rings, Avishai_Two_Rings, Lei_Two_Rings}, and two metallic rings coupled by a p-wave wire was studied in~\cite{David_pWave_Rings}. All these studies focused on equilibrium properties, especially persistent currents inside the rings.

We first study the probability, $P_e$ and $P_h$, for the tunneling electron/hole to flow out from the second lead when an electron is incident with a specific energy from the first lead. The probabilities are, of course, affected by the magnetic fluxes $\Phi_M^{(1)}$ and $\Phi_M^{(2)}$, and we prefer to express them as functions of the magnetic phases $\phi_M^{(1)}$ and $\phi_M^{(2)}$. To solve the problem, we need to combine the scattering process at the tri-junction of the first lead and the left ring, the motion of electron/hole in the left ring, the CAR between the left ring and the right ring, the motion of electron/hole in the right ring and the tunneling process between the right ring and the second lead. The details of the calculation for $P_e$ and $P_h$ are given in the supplementary material~\cite{Supplement_Material}. 

The result of a numerical calculation is shown in Fig.~\ref{fig:Transmission}, under the setting of a group of typical values of the relevant parameters. The magnetic fluxes are varied such that the magnetic phases scan from $0$ to $2\pi$. (Any other situation is equivalent to the one in this range due to the Byers-Yang theorem~\cite{Byers_Yang}.) The electron incident energy $E$ and the MZM width (due to their couplings to the rings) $\Gamma_i=t_i^2/\hslash v_F$ are both set to be much smaller than the MZM coupling energy $E_M$, so that CAR dominates LAR. The circumferences of the two rings are chosen a little bit different in order to distinguish their resonance conditions.

\begin{figure} [t]
\begin{center}
\includegraphics[scale=0.4]{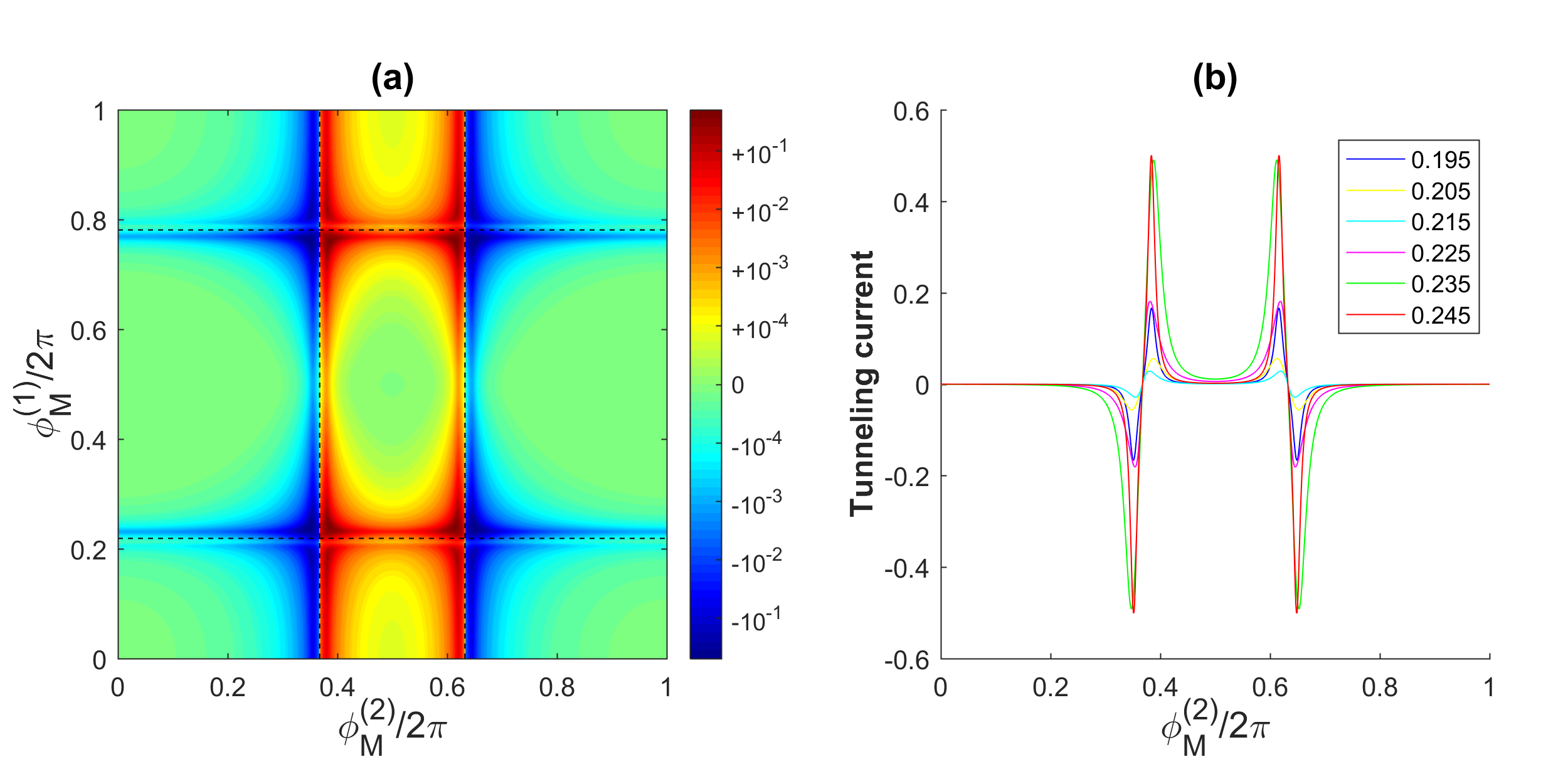}
\end{center}
\caption{Tunneling current in the second lead.  (a) The general picture in terms of the heat map. Dashed lines in the graph correspond to the magnetic fluxes that fulfill $k_F L_1 \pm \phi_M^{(1)} = 2n\pi$ and $k_F L_2 \pm \phi_M^{(2)} = 2n^\prime \pi$, which are $\phi_M^{(1)}=0.219 \times 2\pi, 0.781 \times 2\pi$ and $\phi_M^{(2)}=0.368 \times 2\pi, 0.632 \times 2\pi$ under our setting. (b) The situation when $\phi_M^{(1)}/2\pi = 0.195,\, 0.205,\, 0.215,\, 0.225,\, 0.235,\, 0.245$ is fixed. The current shown in this graph is the relative value $j/j_{\text{(in)}}$. All the parameters are the same as those in Fig. \ref{fig:Transmission}. } \label{fig:Tunnelling_Current}
\end{figure}

From Fig.~\ref{fig:Transmission}, it is clear that the electron/hole tunneling probability peaks when both the resonance conditions $\phi_{e/h}^{(1)}\pm\phi_M^{(1)}=2n\pi$ and $\phi_{e/h}^{(2)}\pm\phi_M^{(2)}=2n^{\prime}\pi$ are fulfilled. Under our setting, $\phi_e$ and $\phi_h$ are only slightly different. So the electron and hole tunneling peaks are very close to each other. Nonetheless, they can be clearly distinguished. The exact locations of the peaks are actually a little further apart than those the resonance condition predicts, due to the quantum mechanical level repulsion.

Next, we consider the tunneling current \cite{Andreev_Scattering_Current}
\begin{equation} \label{eq:Tunnelling_Current}
j=j_{\text{(in)}}[P_e - (v_h/v_e) P_h] \ ,
\end{equation} 
where $j_\text{(in)}$ represents the current carried by the incident electron, $P_{e/h}$ is the electron/hole tunneling probability and $v_{e/h}$ is the velocity. In our case since the incident energy $E$ is small (compared to $E_F$), $v_e \approx v_h \approx v_F$, and so $j \approx j_{\text{(in)}}(P_e - P_h)$. In our model, this current is contributed solely from the MZM-induced CAR process. When $P_e = P_h$, the MZM-induced CAR process does not generate a current, which is the traditional belief and the case for a setup where the tunneling is measured directly across the topological superconducting wire.

Figure~\ref{fig:Tunnelling_Current} shows the tunneling current in the second lead. From Fig.~\ref{fig:Tunnelling_Current}(a), there exist separate regions of positive and negative currents. The positive current corresponds to the situation in which the electron tunneling dominates, while the negative current corresponds to that in which the hole tunneling dominates. The boundaries between these regions are the lines on which $\phi_M^{(2)}$ fulfills $k_F L_2\pm \phi_M^{(2)} = 2 n^\prime \pi$. If we fix the magnetic flux threading the left ring, and vary the flux of the right ring, we see a negative current peak immediately followed by a positive one, or vice versa, as shown in Fig.~\ref{fig:Tunnelling_Current}(b). The separation of the adjacent opposite peaks is due to slightly different wave vectors of electrons and holes. Moreover, the almost equal amplitudes of the peaks imply the tunneling is through MZM, which couple in the same way to electrons as to holes.

The above discussions are restricted to the situation in which a single electron is incident with a specific energy. In realistic situations, we should connect the first lead to a source and the second lead to a drain, and apply a negative voltage bias between them. We should also ground the superconductor hosting the MZM. Electrons are incident from the source, and it depends on the flux threading the second ring whether it is electron or hole that dominates the flow to the drain. When electrons dominate, there is little current flowing through the superconductor to the ground. While if holes dominate, there is an electron current flowing through the superconductor to the ground. The external connection of our set-up and the overall physical phenomenon are analogous to those studied in \cite{DF_MF_Converter}.

Assuming the temperature is well below the bias voltage (and also $E_M$), all the electron states within the range of the bias participate in the transport process, according to the Landauer-B\"uttiker formalism~\cite{Landauer_Buttiker}. This complication, however, does not change our established picture of the tunneling current. The reason is twofold. First, the resonance has a width, originating from the coupling of the ring to the lead and MZM. [This width also determines the phase resolution of a typical tunneling signal shown in Fig.~\ref{fig:Tunnelling_Current}(b)]. If the applied bias is made to be smaller than the resonance width, it is possible that all incident electrons are effectively in resonant states. Second, the positions of tunneling peaks (in the magnetic-phase vs tunneling-current diagram) may slightly vary for different incident electrons, but the overall effect is not that their tunneling currents are canceled, for the wave vectors of the holes are always smaller than those of the electrons, resulting in all the positive current peaks are separated from all the negative ones by a boundary set by the Fermi vector $k_F$. Therefore, the line shape of the tunneling current as shown in Fig.~\ref{fig:Tunnelling_Current}(b) can be qualitatively taken as a signature of MZM-induced CAR.

In our model we implicitly assume that the electrons are spinless, by considering that the electron spin, in an experimental set-up, can be fully polarized in the presence of a Zeeman magnetic field. We also implicitly assume that there is only a single channel in the metallic ring. Actually, we do not intend the rings to have multiple channels, for in that case electrons in different channels can have distinct wave vectors, such that in one magnetic-phase period (from $0$ to $2\pi$) there can exist multiple resonant states, of which if they have overlaps, the tunneling currents may cancel.

Finally, we discuss the parameters used in our numerical calculations and the feasibility of an experimental verification. For the MZM, we believe \cite{Mourik_Ex_MF_Semi_SC_Wire, Deng_Ex_ZBP} that a superconducting gap $\Delta$ of 0.25 meV can be induced in the InSb nanowire by superconducting Nb. When the spin-orbit energy of InSb is 0.3 meV, the coherence length $\xi$ on the wire is estimated to be 185 nm. Then, in order to let the MZM coupling energy $E_M$ ($\approx\Delta e^{-l/\xi}$) to be around 0.01 meV, the length of the InSb nanowire $l$ should be about 595 nm. For the metallic rings, we want the size of the rings to be as large as possible. We follow \cite{GaAs_GaAlAs_Loop} that electron gas at a GaAs-GaAlAs hetero-interface can have a Fermi wavelength $\lambda_F = 42~\text{nm}$ and a Fermi velocity $v_F = 2.6\times 10^5 ~\text{m/s}$, which lead to the Fermi vector $k_F = 0.15 ~\text{nm}^{-1}$ and the Fermi energy $E_F = 12.835~ \text{meV}$. The mobility of the electron gas attains $1.14\times 10^6 ~\text{cm}^2 /\text{Vs}$, which makes the elastic mean free path $11 ~\mu\text{m}$~\cite{GaAs_GaAlAs_review}. Under this condition, we can safely stay in the phase coherent transport regime as we restrict the circumferences of the rings to be no more than 3 $\mu\text{m}$. 

We conclude by noting that the persistent current \cite{Buttiker_Persistent_Current} inside the right ring also changes sign from the tunneling electron dominated regime to the tunneling hole dominated one when the magnetic flux is adjusted. This change in sign of the persistent current can be detected by placing a high sensitivity magnetometer near the right ring. In summary, we have presented a method using the constructive coherence condition in a metallic ring to separate the electron and hole tunneling signals of the Majorana zero mode induced crossed Andreev reflection.

\begin{acknowledgments}
L.F. would like to thank Javad Shabani for helpful discussions on the experimental possibilities. This work was supported by the U.S. DOE BES Core Program (E3B7), and in part by the Center for Integrated Nanotechnologies, a U.S. DOE BES user facility.
\end{acknowledgments}

\end{document}




\title{Tunneling Current Measurement Scheme to Detect 
 Majorana-Zero-Mode-Induced Crossed Andreev Reflection  
  ---  Supplementary Material}

\author{Lei Fang$^{1,2}$}
\email{lfang@gradcenter.cuny.edu} 
\author{David Schmeltzer$^{1,2}$}
\author{Jian-Xin Zhu$^{3,4}$}
\author{Avadh Saxena$^{3}$}
\affiliation{$^{1}$Physics Department, City College of the CUNY, New York, New York 10031, USA} 
\affiliation{$^{2}$Graduate Center, CUNY, 365 5th Ave., New York, New York 10016, USA} 
\affiliation{$^{3}$Theoretical Division, Los Alamos National Laboratory, Los Alamos, New Mexico 87545, USA}
\affiliation{$^{4}$Center for Integrated Nanotechnologies, Los Alamos National Laboratory, Los Alamos, New Mexico 87545, USA}

\maketitle

This supplementary material has three parts. First, we provide a derivation to the scattering matrix of the crossed Andreev reflection (CAR) given in the main text. Next, we study in detail the tunneling process in the system comprising two metallic rings coupled by a pair of overlapped Majorana zero modes (MZM).  Last, we give a physical interpretation of the MZM-induced CAR.

\section{Scattering Matrix Of The Crossed Andreev Reflection}

\begin{figure}  [t]
\begin{center}
\includegraphics[scale=0.45]{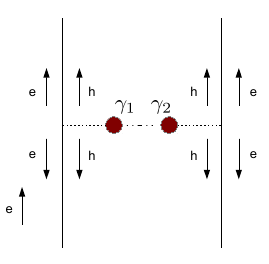}
\end{center}
\caption{Majorana-zero-modes-induced crossed Andreev reflection. An electron is incident from the lower side of the left metallic wire, which is coupled through a pair of Majorana zero modes to the right metallic wire. It can: (1) reflect with amplitude $r_1^{(e)}$; (2) transmit to the upper side with amplitude $t_1^{(e)}$; (3),(4) tunnel to either lower or upper side of the right wire with amplitude $w_1^{(e)}$; (5),(6) be converted to holes through local Andreev reflection and flow to either lower side with amplitude $r_1^{(h)}$ or upper side with amplitude $t_1^{(h)}$; (7),(8) be converted to holes through crossed Andreev reflection and flow to either lower or upper side of the right wire with amplitude $w_1^{(h)}$.} \label{fig:CAR}
\end{figure}

Here we calculate the MZM-induced CAR scattering matrix. As shown in Fig. \ref{fig:CAR}, we consider two metallic wires coupled through a pair of MZM. The Hamiltonian can be written as
\begin{eqnarray}
H \ & = & \, \int dx \ \psi_1^\dagger(x) \big[ \frac{p_1^2}{2m^*} - E_F \big] \psi_1 (x) \nonumber \\ [5pt]
   & + & \  \int dy \ \psi_2^\dagger(y) \big[ \frac{p_2^2}{2m^*} - E_F \big] \psi_2 (y)  \nonumber \\ [5pt]
 & + & \ \int dx \ \frac{t_1}{\sqrt{2}} \gamma_1 \big[ \psi_1^\dagger(x) - \psi_1(x)\big] \delta(x) \nonumber \\ [5pt]
  & + & \ \int dy \ \frac{t_2}{\sqrt{2}} \gamma_2 \big[ \psi_2^\dagger(y) - \psi_2(y)\big] \delta(y) \nonumber \\ [5pt]
  & + & \ i \frac{E_M}{2}\gamma_1 \gamma_2  \ ,   \label{eq:CAR_Hamiltonian}
\end{eqnarray}
where the first two terms describe the two metallic wires, the next two terms denote the coupling between the wires and the MZMs, and the last term represents the coupling between the two MZMs. The quantities $m^*$ and $E_F$ denote the electron effective mass and the Fermi energy respectively.

\vspace{5pt}

Assume an operator 
\begin{eqnarray} 
\Gamma & = & \int dx \big[ D_1(x)\psi_1(x) + B_1(x)\psi_1^\dagger(x) \big] + \frac{C_1}{\sqrt{2}} \gamma_1   \nonumber  \\ [3pt]
& + & \int dy \big[ D_2(y)\psi_2(y) + B_2(y)\psi_2^\dagger(y) \big] + \frac{C_2}{\sqrt{2}} \gamma_2     \hspace{25pt}  \label{eq:CAR_Eigen_Mode}
\end{eqnarray}
diagonalizes the Hamiltonian, in the sense that
\begin{equation}
[\Gamma,H] = E\Gamma  \ ,  \label{eq:CAR_Commutater}
\end{equation}
where $E$ represents the energy of this eigenstate. Substituting (\ref{eq:CAR_Hamiltonian}) and (\ref{eq:CAR_Eigen_Mode}) into (\ref{eq:CAR_Commutater}) and employing the commutation relations $\{\psi_i(x),\psi_j^\dagger(y)\} = \delta_{ij}\delta(x-y)$, $\{\gamma_i , \gamma_j\} = 2\delta_{ij}$, and $\{\psi_i(x),\psi_j(y)\} = \{\psi_i^\dagger(x),\psi_j^\dagger(y)\} = \{ \psi_i(x),\gamma_j \} = \{ \psi_i^\dagger(x),\gamma_j \} = 0$ to simplify the formula, we obtain the Bogoliubov-de Gennes (BdG) equations

\begin{eqnarray}
 \big[\frac{1}{2m^*}p_x^2 - E_F \big] D_1(x) - t_1 \delta(x) C_1 & = & E D_1(x) \ , \\ [5pt]
-\big[\frac{1}{2m^*}p_x^2 - E_F\big] B_1(x) + t_1 \delta(x) C_1 & = & E B_1(x) \ , \\ [5pt]
 t_1 \big[-D_1(0)+B_1(0)\big] - i E_M C_2 & = & E C_1 \ ,  \\ [5pt]
 \big[\frac{1}{2m^*}p_y^2 - E_F \big] D_2(y) - t_2 \delta(y) C_2 & = & E D_2(y) \ , \\ [5pt]  
-\big[\frac{1}{2m^*}p_y^2 - E_F \big] B_2(y) + t_2 \delta(y) C_2 & = & E B_2(y) \ , \\ [5pt]
 t_2 \big[-D_2(0)+B_2(0)\big] + i E_M C_1 & = & E C_2 \ .     
\end{eqnarray}
 
First, suppose an electron is incident from the lower side of the left wire, as shown in Fig. \ref{fig:CAR}. The eigenstate wave-function can be written as

\begin{eqnarray}
D_1(x)\ & = & \ 
\left\lbrace
\begin{array}{cc}
e^{ik_e x} + r_1^{(e)} e^{-ik_e x} \ ,  &  \  x<0  \  ,  \\ [5pt]
t_1^{(e)} e^{ik_e x}  \ ,  &  \  x>0  \  ,
\end{array}
\right.   \\ [5pt]
B_1(x) \  & = & \
\left\lbrace
\begin{array}{cc}
\hspace{10pt} r_1^{(h)} e^{ik_h x} \quad ,  &  \quad  x<0  \  ,  \\ [5pt]
\hspace{10pt} t_1^{(h)} e^{-ik_h x}  \ ,    &  \quad  x>0  \  ,
\end{array}
\right.   \\ [5pt]
D_2(y) \ & = & \
\left\lbrace
\begin{array}{cc}
\hspace{10pt} w_1^{(e)} e^{-ik_e y}   \   ,     &  \quad  y<0  \  ,  \\ [5pt]
\hspace{10pt} w_1^{(e)} e^{ ik_e y} \quad ,     &  \quad  y>0  \  ,
\end{array}
\right.   \\ [5pt] 
B_2(y) \  & = & \
\left\lbrace
\begin{array}{cc}
\hspace{10pt} w_1^{(h)} e^{ ik_h y} \quad     ,  &  \quad  y<0  \  ,  \\ [5pt]
\hspace{10pt} w_1^{(h)} e^{-ik_h y}   \       ,  &  \quad  y>0  \  .
\end{array}
\right.
\end{eqnarray}
Here $k_e = \sqrt{2m^*(E_F+E)}/\hslash$ and $k_h = \sqrt{2m^*(E_F-E)}/\hslash$ are electron and hole wave vectors, $r_1^{(e)}$ and $t_1^{(e)}$ represent electron reflection and transmission amplitudes, $r_1^{(h)}$ and $t_1^{(h)}$ represent LAR amplitudes, $w_1^{(e)}$ and $w_1^{(h)}$ represent electron and hole tunneling amplitudes to the right wire.

Substituting this wave-function ansatz into the BdG equations, we can solve for all the unknown amplitudes. The results are 

\begin{eqnarray}
r_1^{(e)}   &  = & \  i \frac{t_1^2}{\hslash v_e} \big[ E + i(\frac{t_2^2}{\hslash v_e} + \frac{t_2^2}{\hslash v_h}) \big] \bigg/ Z  \ , \\  [5pt]
t_1^{(e)}  & = & \ 1 + r_1^{(e)}  \ , \\ [5pt]
r_1^{(h)}  & = & \ -i \frac{t_1^2}{\hslash v_h} \big[ E + i (\frac{t_2^2}{\hslash v_e} + \frac{t_2^2}{\hslash v_h}) \big] \bigg/ Z  \ , \\  [5pt]
t_1^{(h)}  & = & \ r_1^{(h)}  \ , \\ [5pt]
w_1^{(e)}  & = & \  -E_M \bigg(\frac{t_1 t_2}{\hslash v_e}\bigg) \bigg/ Z  \ , \\ [5pt]
w_1^{(h)}  & = & \   E_M \bigg(\frac{t_1 t_2}{\hslash v_h}\bigg) \bigg/ Z  \  ,
\end{eqnarray}
where $v_e = \hslash k_e / m^*$ and $v_h = \hslash k_h / m^*$ are electron and hole velocities, and  
\begin{equation}
Z = E_M^2 - \left[ E + i(\frac{t_1^2}{\hslash v_e} + \frac{t_1^2}{\hslash v_h}) \right] \left[ E + i(\frac{t_2^2}{\hslash v_e} + \frac{t_2^2}{\hslash v_h}) \right]   \ .
\end{equation}

\vspace{5pt}

Next, if a hole is incident from the lower side of the left wire, we have the wave-function
\begin{eqnarray}
\tilde{B}_1(x)\ & = & \ 
\left\lbrace
\begin{array}{cc}
e^{-ik_h x} + \tilde{r}_1^{(h)} e^{ik_h x} \ ,  &  \  x<0  \  ,  \\ [5pt]
\tilde{t}_1^{(h)} e^{-ik_h x}  \ ,  &  \  x>0  \  ,
\end{array}
\right.   \\ [5pt]
\tilde{D}_1(x) \  & = & \
\left\lbrace
\begin{array}{cc}
\hspace{14pt} \tilde{r}_1^{(e)} e^{-ik_e x}   \   ,    &  \quad  x<0  \  ,  \\ [5pt]
\hspace{14pt} \tilde{t}_1^{(e)} e^{ ik_e x} \quad ,    &  \quad  x>0  \  ,
\end{array}
\right.   \\ [5pt]
\nonumber \\
\tilde{B}_2(y) \ & = & \
\left\lbrace
\begin{array}{cc}
\hspace{10pt} \tilde{w}_1^{(h)} e^{ ik_h y} \quad   ,     &  \quad  y<0  \  ,  \\ [5pt]
\hspace{10pt} \tilde{w}_1^{(h)} e^{-ik_h y}   \     ,     &  \quad  y>0  \  ,
\end{array}
\right.   \\ [5pt] 
\tilde{D}_2(y) \  & = & \
\left\lbrace
\begin{array}{cc}
\hspace{12pt} \tilde{w}_1^{(e)} e^{-ik_e y}    \      ,  &  \quad  y<0  \  ,  \\ [5pt]
\hspace{12pt} \tilde{w}_1^{(e)} e^{ ik_e y} \quad     ,  &  \quad  y>0  \  .
\end{array}
\right.
\end{eqnarray}
Substituting these into the BdG equations and solving them, we obtain
\begin{eqnarray}
\tilde{r}_1^{(h)}   &  = & \ i \frac{t_1^2}{\hslash v_h} \big[ E + i(\frac{t_2^2}{\hslash v_e} + \frac{t_2^2}{\hslash v_h}) \big]   \bigg/ Z  \ , \\  [5pt]
\tilde{t}_1^{(h)}  & = & \ 1 + \tilde{r}_1^{(h)}  \ , \\ [5pt]
\tilde{r}_1^{(e)}  & = & \ -i \frac{t_1^2}{\hslash v_e} \big[ E + i(\frac{t_2^2}{\hslash v_e} + \frac{t_2^2}{\hslash v_h}) \big] \bigg/ Z  \ , \\  [5pt]
\tilde{t}_1^{(e)}  & = & \ \tilde{r}_1^{(e)}  \ , \\ [5pt]
\tilde{w}_1^{(h)}  & = & \ -E_M \bigg(\frac{t_1 t_2}{\hslash v_h}\bigg)  \bigg/ Z  \ , \\ [5pt]
\tilde{w}_1^{(e)}  & = & \  E_M \bigg(\frac{t_1 t_2}{\hslash v_e}\bigg)  \bigg/ Z  \  .
\end{eqnarray}

Similarly, we can consider the processes that an electron or a hole is incident from the upper side of the left wire, or from the lower side of the right wire, or from the upper side of the right wire. We summarize all the results into the scattering matrix
\begin{equation} \label{eq:CAR_S-Matrix}
S_M = \left(
\begin{array}{cccccccc}
r_1^{(e)} & t_1^{(e)} & w_2^{(e)} & w_2^{(e)} & \tilde{r}_1^{(e)} & \tilde{t}_1^{(e)} & \tilde{w}_2^{(e)} & \tilde{w}_2^{(e)}   \\
t_1^{(e)} & r_1^{(e)} & w_2^{(e)} & w_2^{(e)} & \tilde{t}_1^{(e)} & \tilde{r}_1^{(e)} & \tilde{w}_2^{(e)} & \tilde{w}_2^{(e)}   \\
w_1^{(e)} & w_1^{(e)} & r_2^{(e)} & t_2^{(e)} & \tilde{w}_1^{(e)} & \tilde{w}_1^{(e)} & \tilde{r}_2^{(e)} & \tilde{t}_2^{(e)}   \\
w_1^{(e)} & w_1^{(e)} & t_2^{(e)} & r_2^{(e)} & \tilde{w}_1^{(e)} & \tilde{w}_1^{(e)} & \tilde{t}_2^{(e)} & \tilde{r}_2^{(e)}   \\
r_1^{(h)} & t_1^{(h)} & w_2^{(h)} & w_2^{(h)} & \tilde{r}_1^{(h)} & \tilde{t}_1^{(h)} & \tilde{w}_2^{(h)} & \tilde{w}_2^{(h)}   \\
t_1^{(h)} & r_1^{(h)} & w_2^{(h)} & w_2^{(h)} & \tilde{t}_1^{(h)} & \tilde{r}_1^{(h)} & \tilde{w}_2^{(h)} & \tilde{w}_2^{(h)}   \\
w_1^{(h)} & w_1^{(h)} & r_2^{(h)} & t_2^{(h)} & \tilde{w}_1^{(h)} & \tilde{w}_1^{(h)} & \tilde{r}_2^{(h)} & \tilde{t}_2^{(h)}   \\
w_1^{(h)} & w_1^{(h)} & t_2^{(h)} & r_2^{(h)} & \tilde{w}_1^{(h)} & \tilde{w}_1^{(h)} & \tilde{t}_2^{(h)} & \tilde{r}_2^{(h)}
\end{array}
\right) \  ,
\end{equation} 
where the not yet given matrix elements are 
\begin{eqnarray}
r_2^{(e)}   &  = & \  i\frac{t_2^2}{\hslash v_e} \big[ E + i(\frac{t_1^2}{\hslash v_e} + \frac{t_1^2}{\hslash v_h}) \big] \bigg/ Z  \ , \\  [5pt]
t_2^{(e)}  & = & \ 1 + r_2^{(e)}  \ , \\ [5pt]
r_2^{(h)}  & = & \ -i \frac{t_2^2}{\hslash v_h} \big[ E + i (\frac{t_1^2}{\hslash v_e} + \frac{t_1^2}{\hslash v_h}) \big] \bigg/ Z  \ , \\  [5pt]
t_2^{(h)}  & = & \ r_2^{(h)}  \ , \\ [5pt]
w_2^{(e)}  & = & \   E_M \bigg(\frac{t_1 t_2}{\hslash v_e}\bigg) \bigg/ Z  \ , \\ [5pt]
w_2^{(h)}  & = & \  -E_M \bigg(\frac{t_1 t_2}{\hslash v_h}\bigg) \bigg/ Z  \  ,
\end{eqnarray}
and
\begin{eqnarray}
\tilde{r}_2^{(h)}   &  = & \ i \frac{t_2^2}{\hslash v_h} \big[ E + i(\frac{t_1^2}{\hslash v_e} + \frac{t_1^2}{\hslash v_h}) \big]   \bigg/ Z  \ , \\  [5pt]
\tilde{t}_2^{(h)}  & = & \ 1 + \tilde{r}_2^{(h)}  \ , \\ [5pt]
\tilde{r}_2^{(e)}  & = & \ -i \frac{t_2^2}{\hslash v_e} \big[ E + i(\frac{t_1^2}{\hslash v_e} + \frac{t_1^2}{\hslash v_h}) \big] \bigg/ Z  \ , \\  [5pt]
\tilde{t}_2^{(e)}  & = & \ \tilde{r}_2^{(e)}  \ , \\ [5pt]
\tilde{w}_2^{(h)}  & = & \  E_M \bigg(\frac{t_1 t_2}{\hslash v_h}\bigg)  \bigg/ Z  \ , \\ [5pt]
\tilde{w}_2^{(e)}  & = & \ -E_M \bigg(\frac{t_1 t_2}{\hslash v_e}\bigg)  \bigg/ Z  \  .
\end{eqnarray}

\section{Transport process in the system of two coupled rings }

For simplicity, we suppose that both the electron and the hole tunneling processes at the two tri-junctions can be described by the same scattering matrix
\begin{equation} \label{eq:Tri_Junction_S-Matrix}
S_L = \left(
\begin{array}{ccc}
-(a+b) & \sqrt{\epsilon} & \sqrt{\epsilon} \\
\sqrt{\epsilon} &  a  &  b  \\
\sqrt{\epsilon} &  b  &  a
\end{array}
\right) \, ,
\end{equation}  
where $a=(\sqrt{1-2\epsilon}-1)/2$,  $b=(\sqrt{1-2\epsilon}+1)/2$ and $\epsilon$ denotes the tunneling strength.

\begin{figure}  [t]
\begin{center}
\includegraphics[scale=0.4]{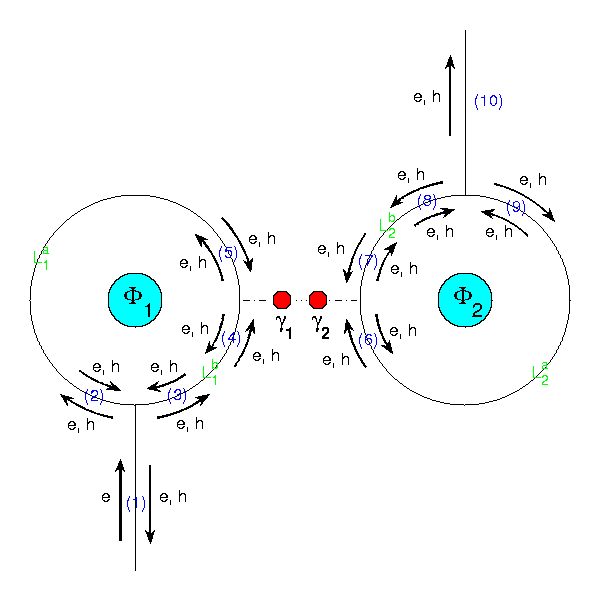}
\end{center}
\caption{The setup to measure MZM induced CAR. $\gamma_1$ and $\gamma_2$ are two MZM that have an overlap. Two metallic rings are coupled to the two MZM separately. The rings are threaded by magnetic fluxes $\Phi_1$ and $\Phi_2$. Two external leads contact the two rings. The electrons are incident from the first lead. The tunneling electrons and the tunneling holes flow out from the second lead.} \label{fig:Transport_Configuration_Details}
\end{figure}

Now we start to set a group of equations to describe the steady transport process. Let $D$ and $B$ represent electron and hole amplitudes respectively. Let the subscripts $1$ and $2$ denote counter-clockwise and clockwise electron/hole motions in the ring respectively, the superscript $(i)$ is the corresponding part as shown in Fig. \ref{fig:Transport_Configuration_Details}. 

Suppose that an electron is incident from the first lead. The tunneling process at the tri-junction between the first lead and the left ring can be described as  
\begin{equation}
S_L \;
\left(
\begin{array}{c}
1 \\ D^{(2)}_1 \\ D^{(3)}_2
\end{array}
\right)
\quad = \quad
\left(
\begin{array}{c}
D^{(1)} \\ D^{(2)}_2 \\ D^{(3)}_1
\end{array}
\right)  \;  ,
\end{equation}
\begin{equation}
S_L \;
\left(
\begin{array}{c}
0 \\ B^{(2)}_1 \\ B^{(3)}_2
\end{array}
\right)
\quad = \quad
\left(
\begin{array}{c}
B^{(1)} \\ B^{(2)}_2 \\ B^{(3)}_1
\end{array}
\right)  \;  ,
\end{equation}
where the electron and hole incoming amplitudes are $1$ and $0$ on the first lead respectively. $D^{(1)}$ and $B^{(1)}$ denote the electron and the hole outgoing amplitudes.

The propagation in the left ring can be described by
\begin{equation}
D^{(5)}_1 = D^{(2)}_1  \exp[i(-k_e L_1^a -\phi_1^a)] \ ,
\end{equation}
\begin{equation}
D^{(5)}_2 = D^{(2)}_2  \exp[i(+k_e L_1^a -\phi_1^a)] \ ,
\end{equation}
\begin{equation}
B^{(5)}_1 = B^{(2)}_1  \exp[i(+k_h L_1^a +\phi_1^a)] \ , 
\end{equation}
\begin{equation}
B^{(5)}_2 = B^{(2)}_2  \exp[i(-k_h L_1^a +\phi_1^a)] \ ,
\end{equation}
\begin{equation}
D^{(4)}_1 = D^{(3)}_1  \exp[i(+k_e L_1^b +\phi_1^b)] \ ,
\end{equation}
\begin{equation}
D^{(4)}_2 = D^{(3)}_2  \exp[i(-k_e L_1^b +\phi_1^b)] \ ,
\end{equation}
\begin{equation}
B^{(4)}_1 = B^{(3)}_1  \exp[i(-k_h L_1^b -\phi_1^b)] \ ,
\end{equation}
\begin{equation}
B^{(4)}_2 = B^{(3)}_2  \exp[i(+k_h L_1^b -\phi_1^b)] \ ,
\end{equation}
where $L_1^a$ ($L_1^b$) denotes the length of the arm $a$ ($b$) of the left ring, and $\phi_1^a$ ($\phi_1^b$) denotes the magnetic phase an electron/hole acquires when circulating counter-clockwise/clockwise along the arm $a$ ($b$). 

The crossed Andreev reflection between the left ring and the right ring can be described as
\begin{equation}
S_M  \;
\left(
\begin{array}{c}
D^{(5)}_2 \\ D^{(4)}_1 \\ D^{(7)}_1 \\ D^{(6)}_2 \\ B^{(5)}_2 \\ B^{(4)}_1 \\ B^{(7)}_1 \\ B^{(6)}_2
\end{array}
\right)
\quad  =  \quad
\left(
\begin{array}{c}
D^{(5)}_1 \\ D^{(4)}_2 \\ D^{(7)}_2 \\ D^{(6)}_1 \\ B^{(5)}_1 \\ B^{(4)}_2 \\ B^{(7)}_2 \\ B^{(6)}_1
\end{array}
\right)
\end{equation}

The propagation in the right ring can be described by
\begin{equation}
D^{(9)}_1 = D^{(6)}_1 \exp[i(+k_e L_2^a +\phi_2^a)]  \  ,
\end{equation}
\begin{equation}
D^{(9)}_2 = D^{(6)}_2 \exp[i(-k_e L_2^a +\phi_2^a)]  \  ,
\end{equation}
\begin{equation}
B^{(9)}_1 = B^{(6)}_1 \exp[i(-k_e L_2^a -\phi_2^a)]  \  ,
\end{equation}
\begin{equation}
B^{(9)}_2 = B^{(6)}_2 \exp[i(+k_e L_2^a -\phi_2^a)]  \  ,
\end{equation}
\begin{equation}
D^{(8)}_1 = D^{(7)}_1 \exp[i(-k_e L_2^b -\phi_2^b)]  \  ,
\end{equation}
\begin{equation}
D^{(8)}_2 = D^{(7)}_2 \exp[i(+k_e L_2^b -\phi_2^b)]  \  ,
\end{equation}
\begin{equation}
B^{(8)}_1 = B^{(7)}_1 \exp[i(+k_e L_2^b +\phi_2^b)]  \  ,
\end{equation}
\begin{equation}
B^{(8)}_2 = B^{(7)}_2 \exp[i(-k_e L_2^b +\phi_2^b)]  \  ,
\end{equation}
where $L_2^a$ ($L_2^b$) denotes the length of the arm $a$ ($b$) of the right ring, and $\phi_2^a$ ($\phi_2^b$) denotes the magnetic phase an electron/hole acquires when circulating counter-clockwise/clockwise along the arm $a$ ($b$). 

The tunneling process at the tri-junction between the right ring and the second lead can be described as
\begin{equation}
S_L \;
\left(
\begin{array}{c}
0 \\ D^{(8)}_2 \\ D^{(9)}_1
\end{array}
\right)
\quad = \quad
\left(
\begin{array}{c}
D^{(10)} \\ D^{(8)}_1 \\ D^{(9)}_2
\end{array}
\right)  \;  ,
\end{equation}
\begin{equation}
S_L \;
\left(
\begin{array}{c}
0 \\ B^{(8)}_2 \\ B^{(9)}_1
\end{array}
\right)
\quad = \quad
\left(
\begin{array}{c}
B^{(10)} \\ B^{(8)}_1 \\ B^{(9)}_2
\end{array}
\right)  \;  ,
\end{equation}
where on the second lead electron and hole incoming amplitudes are both $0$. $D^{(10)}$ and $B^{(10)}$ denote the electron and hole outgoing amplitudes separately.

Combining these equations, we can solve out all the amplitudes. Due to the linearity, they are easy to solve numerically. What we concern most are $D^{(10)}$ and $B^{(10)}$, since $P_e = |D^{(10)}|^2$ and $P_h = |B^{(10)}|^2$ represent the electron and hole tunneling probabilities respectively.

\section{A physical interpretation}

\begin{figure}  [t]
\begin{center}
\includegraphics[scale=0.45]{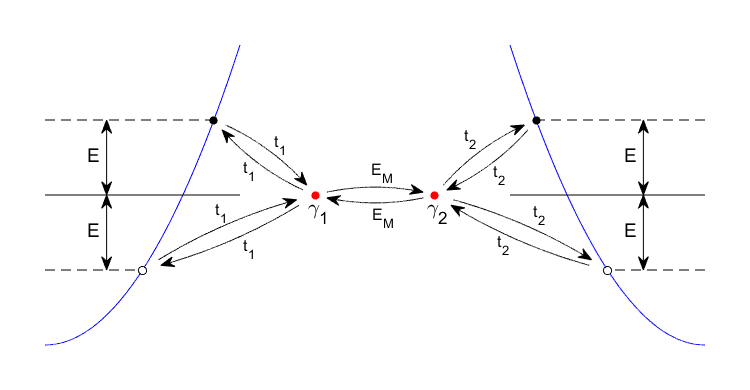}
\end{center}
\caption{ Tunneling processes that consist of  Majorana-zero-modes-induced crossed Andreev reflection. Electrons and holes in two metallic leads couple to the two MZM with amplitudes $t_1$ and $t_2$. The coupling between two MZM can be taken as the possibility of hopping between each other with amplitude $E_M$. } \label{fig:Dispersion}
\end{figure}

Electrons and holes have equal tunneling amplitudes to MZM, due to the special form of the coupling between metallic leads and MZMs. The entire process of MZM-induced CAR is illustrated in Fig. \ref{fig:Dispersion}, from which it is readily seen that electron and hole have the same status as long as only the energy is considered.  

We can also get from this graph the reason why MZM-induced CAR dominates LAR when MZM coupling energy $E_M$ is much larger than both the electron/hole incident energy $E$ and the tunneling amplitude $t_{1,2}$ between metallic leads and MZM. Take the MZM coupling term as a perturbation, and the isolated MZMs have Green's functions of the form 
\begin{equation}
G_i^{(0)}(E) = \frac{1}{E + 2i t_i^2/\hbar v_F}   \;   ,
\end{equation} 
where the subscript $i$ ($=1, 2$) represents the $i$-th MZM. The imaginary part in the denominator of this expression results from the coupling of MZM to the external lead. The CAR is dominated by the propagator from $\gamma_1$ to $\gamma_2$ 
\begin{equation}
G_1^{(0)} E_M G_2^{(0)} + G_1^{(0)} E_M G_2^{(0)} E_M G_1^{(0)} E_M G_2^{(0)} + ...  \;  ,
\end{equation} 
while the LAR is dominated by the propagator from $\gamma_1$ to $\gamma_1$ itself
\begin{equation}
G_1^{(0)} + G_1^{(0)} E_M G_2^{(0)} E_M G_1^{(0)} + ...  \;  .
\end{equation} 
It is obvious that the amplitude of CAR has a factor $E_M/(E+2i t_2^2/\hbar v_F)$ ($= E_M G_2^{(0)}$) larger than the amplitude of LAR.

